\documentclass[twocolumn]{aastex631}

\usepackage{amsmath}

\newcommand{\pivec}{\mbox{\boldmath $\pi$}}
\newcommand{\muvec}{\mbox{\boldmath $\mu$}}

\lefthead{} 
\righthead{}

\begin{document}

\title{
Using Source Proper Motion to Validate Terrestrial Parallax: OGLE-2019-BLG-1058
}

% Author List ------------------------------------------------------------------------------------------------------------
% leading author -----------------------------
\author{In-Gu~Shin} 
%\altaffiliation{The KMTNet Collaboration}
\affiliation{Korea Astronomy and Space Science Institute, Daejon 014025, Republic of Korea}
%\correspondingauthor{In-Gu~Shin} \email{ingushin@gmail.com}
%
\author{Jennifer~C.~Yee}
%\altaffiliation{The KMTNet Collaboration}
\affiliation{Center for Astrophysics $|$ Harvard \& Smithsonian 60 Garden St., Cambridge, MA 02138, USA}
\author{Kyu-Ha~Hwang}
%\altaffiliation{The KMTNet Collaboration}
\affiliation{Korea Astronomy and Space Science Institute, Daejon 014025, Republic of Korea}
\author{Andrzej~Udalski}
%\altaffiliation{The OGLE Collaboration}
\affiliation{Astronomical Observatory, University of Warsaw, Al.~Ujazdowskie 4, 00-478 Warszawa, Poland}
\author{Andrew~Gould}
%\altaffiliation{The KMTNet Collaboration}
\affiliation{Max Planck Institute for Astronomy, K\"onigstuhl 17, D-69117 Heidelberg, Germany}
\affiliation{Department of Astronomy, The Ohio State University, 140 W. 18th Ave., Columbus, OH 43210, USA}
\collaboration{6}{(Leading authors),}
%
% KMTNet ---------------------------
% Science Team
\author{Michael~D.~Albrow} 
\affiliation{University of Canterbury, Department of Physics and Astronomy, Private Bag 4800, Christchurch 8020, New Zealand}
\author{Sun-Ju~Chung}
\affiliation{Korea Astronomy and Space Science Institute, Daejon 014025, Republic of Korea}
\affiliation{University of Science and Technology, Korea, (UST), 217 Gajeong-ro, Yuseong-gu, Daejeon 34113, Republic of Korea}
\author{Cheongho~Han}
\affiliation{Department of Physics, Chungbuk National University, Cheongju 28644, Republic of Korea}
\author{Youn~Kil~Jung}
\affiliation{Korea Astronomy and Space Science Institute, Daejon 014025, Republic of Korea}
\affiliation{University of Science and Technology, Korea, (UST), 217 Gajeong-ro, Yuseong-gu, Daejeon 34113, Republic of Korea}
\author{Hyoun-Woo~Kim}
\affiliation{Korea Astronomy and Space Science Institute, Daejon 014025, Republic of Korea}
\author{Yoon-Hyun~Ryu}
\affiliation{Korea Astronomy and Space Science Institute, Daejon 014025, Republic of Korea}
\author{Yossi~Shvartzvald}
\affiliation{Department of Particle Physics and Astrophysics, Weizmann Institute of Science, Rehovot 76100, Israel}
\author{Weicheng~Zang}
\affiliation{Department of Astronomy and Tsinghua Centre for Astrophysics, Tsinghua University, Beijing 100084, China}
%
% Operation Team
\author{Sang-Mok~Cha}
\affiliation{Korea Astronomy and Space Science Institute, Daejon 014025, Republic of Korea}
\affiliation{School of Space Research, Kyung Hee University, Yongin, Kyeonggi 17104, Republic of Korea}
\author{Dong-Jin~Kim}
\affiliation{Korea Astronomy and Space Science Institute, Daejon 014025, Republic of Korea}
\author{Seung-Lee~Kim} 
\affiliation{Korea Astronomy and Space Science Institute, Daejon 014025, Republic of Korea}
\affiliation{University of Science and Technology, Korea, (UST), 217 Gajeong-ro, Yuseong-gu, Daejeon 34113, Republic of Korea}
\author{Chung-Uk~Lee}
\affiliation{Korea Astronomy and Space Science Institute, Daejon 014025, Republic of Korea}
\author{Dong-Joo~Lee}
\affiliation{Korea Astronomy and Space Science Institute, Daejon 014025, Republic of Korea}
\author{Yongseok~Lee}
\affiliation{Korea Astronomy and Space Science Institute, Daejon 014025, Republic of Korea}
\affiliation{School of Space Research, Kyung Hee University, Yongin, Kyeonggi 17104, Republic of Korea}
\author{Byeong-Gon~Park}
\affiliation{Korea Astronomy and Space Science Institute, Daejon 014025, Republic of Korea}
\affiliation{University of Science and Technology, Korea, (UST), 217 Gajeong-ro, Yuseong-gu, Daejeon 34113, Republic of Korea}
\author{Richard~W.~Pogge}
\affiliation{Department of Astronomy, The Ohio State University, 140 W. 18th Ave., Columbus, OH 43210, USA}
\collaboration{17}{(The KMTNet Collaboration),}
%
% OGLE -----------------------------
\author{Przemek~Mr{\'o}z}
\affiliation{Astronomical Observatory, University of Warsaw, Al.~Ujazdowskie 4, 00-478 Warszawa, Poland}
\affiliation{Division of Physics, Mathematics, and Astronomy, California Institute of Technology, Pasadena, CA 91125, USA}
\author{Micha{\l}~K.~Szyma{\'n}ski}
\affiliation{Astronomical Observatory, University of Warsaw, Al.~Ujazdowskie 4, 00-478 Warszawa, Poland}
\author{Jan~Skowron}
\affiliation{Astronomical Observatory, University of Warsaw, Al.~Ujazdowskie 4, 00-478 Warszawa, Poland}
\author{Radek~Poleski}
\affiliation{Astronomical Observatory, University of Warsaw, Al.~Ujazdowskie 4, 00-478 Warszawa, Poland}
\author{Igor~Soszy{\'n}ski}
\affiliation{Astronomical Observatory, University of Warsaw, Al.~Ujazdowskie 4, 00-478 Warszawa, Poland}
\author{Pawe{\l}~Pietrukowicz}
\affiliation{Astronomical Observatory, University of Warsaw, Al.~Ujazdowskie 4, 00-478 Warszawa, Poland}
\author{Szymon~Koz{\l}owski}
\affiliation{Astronomical Observatory, University of Warsaw, Al.~Ujazdowskie 4, 00-478 Warszawa, Poland}
\author{Krzysztof~Ulaczyk}
\affiliation{Department of Physics, University of Warwick, Gibbet Hill Road, Coventry, CV4 7AL, UK}
\author{Krzysztof~A.~Rybicki}
\affiliation{Astronomical Observatory, University of Warsaw, Al.~Ujazdowskie 4, 00-478 Warszawa, Poland}
\author{Patryk~Iwanek}
\affiliation{Astronomical Observatory, University of Warsaw, Al.~Ujazdowskie 4, 00-478 Warszawa, Poland}
\author{Marcin~Wrona}
\affiliation{Astronomical Observatory, University of Warsaw, Al.~Ujazdowskie 4, 00-478 Warszawa, Poland}
\author{Mariusz~Gromadzki}
\affiliation{Astronomical Observatory, University of Warsaw, Al.~Ujazdowskie 4, 00-478 Warszawa, Poland}
\collaboration{12}{(The OGLE Collaboration)}    
% ------------------------------------------------------------------------------------------------------------------------

% 0. ABSTRACT -----------------------------------------------------------------------------------------
\begin{abstract}
We show that because the conditions for producing terrestrial microlens parallax (TPRX; i.e., a nearby disk lens) will also tend to produce a large lens-source relative proper motion ($\mu_{\rm rel}$), source proper motion ($\muvec_{\rm S}$) measurements in general provide a strong test of TPRX signals, which \citet{gould13} showed were an important probe of free-floating planet (FFP) candidates. As a case study, we report a single-lens/single-source microlensing event designated as OGLE-2019-BLG-1058. For this event, the short timescale ($\sim 2.5$ days) and very fast $\mu_{\rm rel}$ ($\sim 17.6\, {\rm mas\, yr^{-1}}$) suggest that this isolated lens is a FFP candidate located in the disk of our Galaxy. For this event, we find a TPRX signal consistent with a disk FFP, but at low significance. A direct measurement of the $\muvec_{\rm S}$ shows that the large $\mu_{\rm rel}$ is due to an extreme $\muvec_{\rm S}$, and thus, the lens is consistent with being a very low-mass star in the bulge and the TPRX measurement is likely spurious. By contrast, we show how a precise measurement of $\muvec_{\rm S}$ with the mean properties of the bulge proper motion distribution would have given the opposite result; i.e., provided supporting evidence for an FFP in the disk and the TPRX measurement.
\end{abstract}
\keywords{Gravitational microlensing (672) --- Microlensing parallax (2144) --- Gravitational microlensing exoplanet detection (2147) --- Free floating planets (549)}

% 1. INTRODUCTION -------------------------------------------------------------------------------------
\section{Introduction} \label{ss1}

Various scenarios predict the existence of free-floating planets (hereafter, FFPs). For example, FFPs can be planets that are unbound from their host stars by various dynamical mechanisms such as planet-planet scattering \citep{rasio96, weidenschilling96, chatterjee08}, ejections from the multiple-star system \citep{kaib13} or stellar clusters \citep{spurzem09}, stellar flyby \citep{malmberg11}, or post-main-sequence evolution of the host star(s) \citep{veras11}. Alternatively, other FFP formation scenarios posit that FFPs can be formed similar to stars, i.e., formed from the fragmentation of gas clouds \citep{boyd05, whitworth06} or cloudlets in HII regions \citep{gahm07, grenman14}. A small number of FFPs (or FFP candidates) have been discovered through direct detections of their flux \citep{bardalez20, delorme12, gagne14a, gagne14b, gagne15, gagne17, kellogg16, marsh10, liu13, luhman05, schneider14, schneider16}. These detections tend to be very young objects with masses several times that of Jupiter.

Indeed, discovering FFPs is challenging because of the faintness of the object and lack of interaction with a host star (i.e., the host does not exist by definition). The microlensing method can detect any objects based on an astronomical phenomenon, which is that the brightness of the background object (i.e., source) is magnified when the foreground object (i.e., lens) approaches/crosses the line of sight between the observer and the source. Thus, the method does not rely on the brightness of the lens (e.g., FFPs). Also, the method is independent of the interaction with the host star. As a result, in principle, the microlensing method can be a robust method to discover FFPs. Moreover, the method can provide systematic FFP samples in our Galaxy.

\citet{sumi11} presented the first statistical study of FFP candidates with microlensing. Based on a statistical excess of short timescale events at $t_{\rm E} \sim 1$ day, they suggested that Jupiter-mass FFPs are more common than stars in our Galaxy. However, a later analysis from the Optical Gravitational Lensing Experiment \citep[OGLE:][]{udalski15} survey was inconsistent with such an excess, although it did show a population of events with even shorter timescales, which could be explained by super-Earth FFPs \citep{mroz17}.  

The microlensing event produced by an FFP lasts a relatively short time because of the low-mass lens. The timescale of the event ($t_{\rm E}$) is defined as
\begin{equation}
t_{\rm E} \equiv \frac{\theta_{\rm E}}{\mu_{\rm rel}} ~~;~~ \theta_{\rm E} = \sqrt{\kappa M_{\rm L} \pi_{\rm rel}}
\label{eqn:tE_thetaE}
\end{equation}
where the $\theta_{\rm E}$ and $\mu_{\rm rel}$ are the angular Einstein ring radius and the lens-source relative proper motion, respectively. The $\theta_{\rm E}$ can be defined using the lens mass ($M_{\rm L}$) and the lens-source relative parallax ($\pi_{\rm rel}$). The relative parallax is defined as $\pi_{\rm rel} \equiv {\rm au} / D_{\rm rel}$, where $D_{\rm rel}$ is relative distance to the lens ($D_{\rm L}$) and source ($D_{\rm S}$), which is defined as $D_{\rm rel} \equiv (D_{\rm L}^{-1} - D_{\rm S}^{-1})^{-1}$. Finally, $\kappa$ is a constant defined as $\kappa = 4 G/c^2 {\rm au} = 8.144\, {\rm mas}\, M_{\odot}^{-1}$.

If both the source and lens are located in the bulge (which is a typical case for microlensing events), we can estimate 
\begin{equation}
t_{\rm E} \approx 0.7\ \mathrm{days} \left(\frac{M_{\rm L}}{M_{\rm Jup}}\right)^{1/2} \left(\frac{\pi_{\rm rel}}{16\, \mu{\rm as}}\right)^{1/2} \left(\frac{\mu_{\rm rel}}{5.8\, {\rm mas}\, {\rm yr}^{-1}}\right). 
\label{eqn:tE_estimation}
\end{equation}
This event timescale is very short compared to the case of a stellar-mass lens with $M_{\rm L} \sim 0.5\, M_{\odot} \sim 520\, M_{\rm Jup}$ would give $t_{\rm E} \sim 16.1$ days. Thus, to capture these short timescale events, continuous observations with high-cadence are required. 

Despite the discovery of short timescale microlensing events, it is another story to determine/confirm the properties of discovered FFPs. The lens mass ($M_{\rm L}$) and the distance to the lens ($D_{\rm L}$) can be determined as 
\begin{equation}
M_{\rm L} = \frac{\theta_{\rm E}}{\kappa \pi_{\rm E}} ~~;~~ 
D_{\rm L} = \frac{\rm AU}{\pi_{\rm E}\theta_{\rm E} + \pi_{\rm S}},
\label{eqn:ML_DL}
\end{equation}
where $\pi_{\rm E}$ is the amplitude of the microlens parallax vector. The parallax of the source is defined as $\pi_{\rm S} \equiv {\rm AU}/D_{\rm S}$. Thus, to obtain conclusive lens properties, two observables are required: the angular Einstein ring radius ($\theta_{\rm E}$) and the microlens parallax ($\pivec_{\rm E}$). 

The Einstein radius, $\theta_{\rm E}$, is a fundamental unit for the analysis of the microlensing event. It can be determined when the finite-source effect is detected. The effect is caused by the finite size of the source, which affects the magnification pattern of the light curve, i.e., the magnification becomes moderate. The angular source radius ($\theta_{\ast}$) can be parameterized by normalizing it to the angular Einstein ring radius ($\theta_{\rm E}$). That is, $\rho_{\ast} \equiv \theta_{\ast} / \theta_{\rm E}$. The observable $\theta_{\ast}$ can be measured with multi-band observations using the method described in \citet{yoo04}.

Recently, the OGLE and Korea Microlensing Telescope Network \citep[KMTNet:][]{kim16} surveys have detected several strong FFP candidates with detections of finite source effects \citep[e.g., see][]{mroz18, mroz19, mroz20a, mroz20b}. The KMTNet microlensing survey is particularly useful for FFP studies because of its continuous high-cadence observations, which enable both detection of short-timescale events and characterization of finite source effects. Indeed, there have also been some FFP candidate detections from the KMTNet survey alone \citep[i.e.,][]{kimhw21, ryu21}. Thus, FFP candidates are now being routinely discovered by microlensing.

In fact, \citet{ryu21} have described initial work on constructing a $\theta_{\rm E}$ distribution to better constrain the FFP population. They posit that there may exist an ``Einstein desert" in the distribution of $\theta_{\rm E}$, such that events with $\theta_{\rm E} \lesssim 10\, \mu$as reflect a distinct population of FFPs. However, even with such a distinction between the population of ``stellar/BD" lenses and ``FFP" lenses, because the underlying distributions are continuous, this ``desert" is really a minimum, and FFP lenses with larger $\theta_{\rm E}$ can exist. At the same time, many of the lenses at larger $\theta_{\rm E}$ will be stars rather than FFPs. The trick is to distinguish between them.

\citet{gould13} showed that the terrestrial microlens parallax effect can be used to probe the FFP and brown dwarf populations by making direct mass measurements for individual lenses (that also have finite source effects). In terrestrial microlens parallax \citep[TPRX:][]{gould97}, $\pi_{\rm E}$ can be measured with simultaneous observations from observatories located on different continents. \citet{gould13} derive an analytic condition for detecting the TPRX signal: $\rho_{\ast} \tilde{r}_{\rm E} \lesssim 50\, R_{\oplus}$, where $\tilde{r}_{\rm E} \equiv \mathrm{au}/\pi_{\rm E}$ is the Einstein ring radius projected on the observer plane. The condition can be rewritten as $\theta_{\ast} \left( D_{\rm L}^{-1} - D_{\rm S}^{-1} \right)^{-1} \lesssim 50\, R_{\oplus}$. For a typical source that lies in the bulge, i.e., $\theta_{\ast} \sim 0.6\, \mu{\rm as}$ and $D_{\rm S} \sim 8$ kpc, the distance to the lens is a key factor for satisfying the condition. It implies that the best chance to detect TPRX would be a disk-lens event ($D_{\rm L} \lesssim 2.5$ kpc). 

Such a nearby disk-lens is likely to have a high proper motion. As a result, we can expect the microlensing event produced by such a lens to have a fast lens-source relative proper motion. Thus, the $\mu_{\rm rel}$ is an efficient indicator to estimate the location of the lens because $\mu_{\rm rel}$ ($= \theta_{\ast} / \rho_{\ast} t_{\rm E}$) can be routinely determined for finite source events. In practice, even if the finite-source effect is weak, (at least) the maximum $\rho_{\ast}$ value can be determined. It yields a minimum $\mu_{\rm rel}$ value. A typical bulge-lens/bulge-source event shows $\mu_{\rm rel} = 5 \sim 10\, {\rm mas\, yr^{-1}}$, so $\mu_{\rm rel} \gtrsim 10\, {\rm mas\, yr^{-1}}$ can provide evidence in favor of a lens in the disk. 

However, a large $\mu_{\rm rel}$ can also be produced by a lens and a source in the bulge but at opposite extremes of the velocity distribution. While such a pair of velocities is unlikely, it is countered by a large increase in volume for bulge lenses as compared to lenses in the nearby disk. In addition, low-mass stellar lenses in the bulge (i.e., with very small $\pi_{\rm rel}$) are expected to be the primary contaminants for FFP candidates that have $\theta_{\rm E} \gtrsim 10\ \mu$as. At the same time, many FFP candidates have bright sources \citep[e.g., see discussion in ][]{ryu21}, so the source proper motion $\muvec_{\rm S}$ can often be directly measured. This measurement can then be compared to the proper motion distribution of bulge stars to check whether the lens could plausibly be a member of the bulge population given $\mu_{\rm rel}$ and $\muvec_{\rm S}$.

\citet{bennett14} applied this test in the case of MOA-2011-BLG-262 for which they found a source proper motion $\muvec_{\rm S,hel}(b, \ell) = (-2.3, -0.9) \pm (2.8, 2.6)\, {\rm mas\, yr^{-1}}$. In the case of their ``fast" solution with $\mu_{\rm rel} \sim 19.6\, {\rm mas\, yr^{-1}}$, the source proper motion measurement gave roughly equal probability to disk lenses as bulge lenses in a Bayesian analysis. By contrast, for their ``slow" solution with $\mu_{\rm rel} \sim 11.6\, {\rm mas\, yr^{-1}}$, bulge lenses are strongly favored, as would be expected given the much larger volume for distant lenses and the relative lack of tension with two stars drawn from a distribution with $\sigma_{\mu} \sim 3\, {\rm mas\, yr^{-1}}$ in each direction.

As another example, \citet{mroz20b} found $\mu_{\rm rel} = 10.6\, {\rm mas\, yr^{-1}}$ for OGLE-2016-BLG-1928, but a source proper motion very close to the mean of the proper motion distribution for bulge giants (i.e., $\mu_{\rm rel} = 5 \sim 10\, {\rm mas\, yr^{-1}}$). Because of the source proper motion, they argue that the lens is more consistent with being a member of the disk. These two events demonstrate that if the source proper motion is the same as or typical of the mean bulge, a large $\mu_{\rm rel}$ can strongly favor a disk lens, and the larger the $\mu_{\rm rel}$, the less close $\muvec_{\rm S}$ needs to be to the mean.

Hence, because FFP events with TPRX are expected to be in the near disk and, therefore, have large $\mu_{\rm rel}$, measuring the source proper motion can be a test for validating the TPRX measurement. In this paper, we present an example of such a test. We report the analysis of OGLE-2019-BLG-1058, which shows a very short timescale, $t_{\rm E} \sim 2.5$ days. Also, there exists a clear signature of the finite-source effect in the peak of the light curve. Thus, $\theta_{\rm E}$ and $\mu_{\rm rel}$ can be determined. The $\mu_{\rm rel}$ of the event is very fast ($\mu_{\rm rel} \sim 17.5 \,{\rm mas\, yr^{-1}}$), which implies that the lens is likely to be located in the disk. This in turn suggests that we can find the signal of the TPRX effect for this event. In Section \ref{ss2}, we present observations of this event. In Section \ref{ss3}, we describe the analyses of the light curve including the $\theta_{\rm E}$ determination and the TPRX measurement. In Section \ref{ss4}, we present the lens properties, as well as the proper motion of the source. Also, we check for the existence of a more-massive host star in Section \ref{ss5}. Lastly, we present our conclusions and discuss our findings in Section \ref{ss6}. 

% Figure 1 -----------------------------------------------------------------------------------------
\begin{figure*}[htb!]
\epsscale{1.00}
\plotone{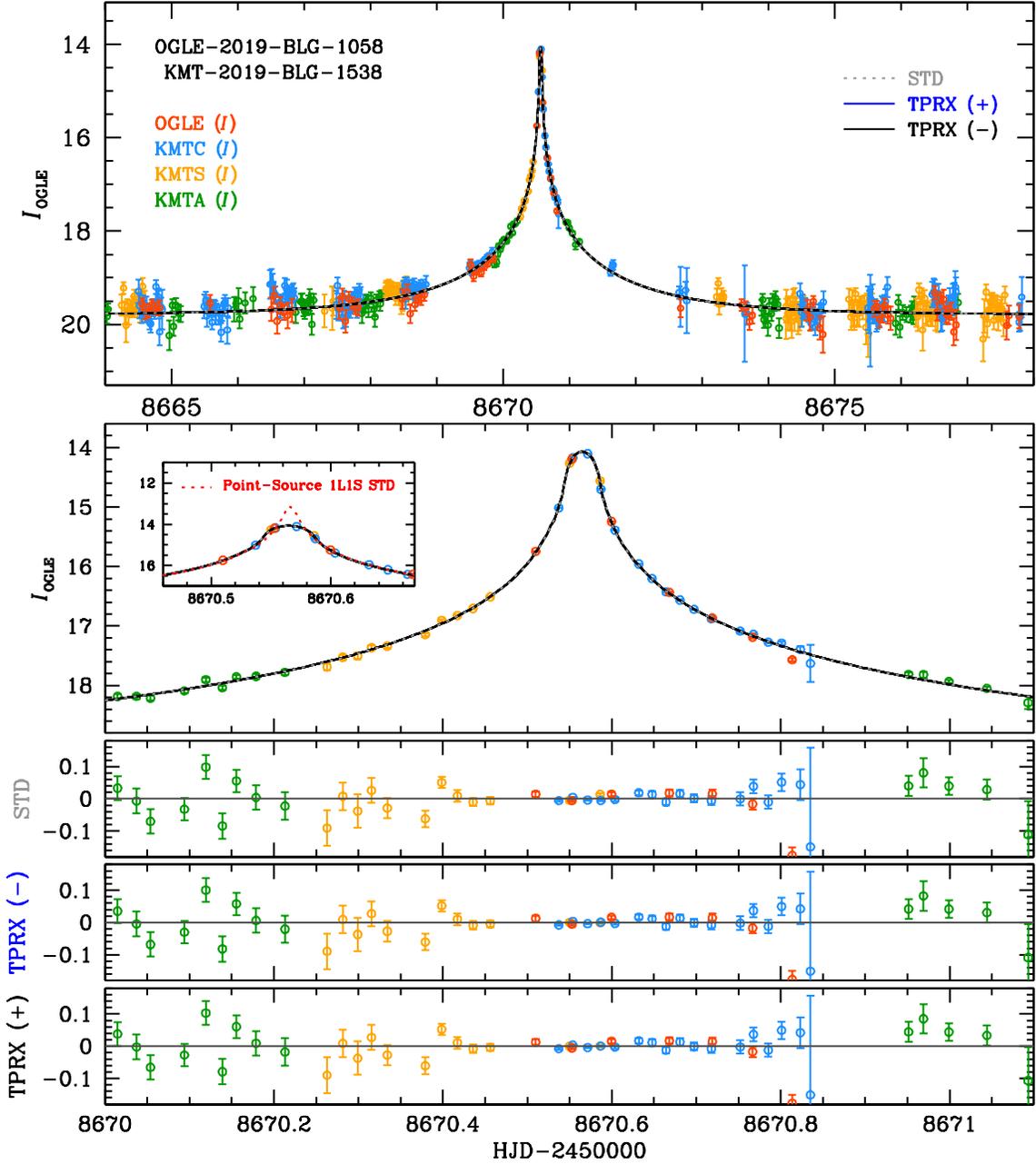}
\caption{
The observed light curve of OGLE-2019-BLG-1058 with model light curves. The red 
dashed line indicates the best-fit static model (STD). The blue and black lines indicate the best-fit 
models including the terrestrial microlens parallax (TPRX) of $(u_0 > 0; +)$ and $(u_0 < 0; -)$ cases, 
respectively. Three bottom panels show the residual between each model and observations.
\label{fig:LCs}}
\end{figure*}
% --------------------------------------------------------------------------------------------------

\section{Observations} \label{ss2}

A microlensing event designated OGLE-2019-BLG-1058 occurred on a background star (source) located at $(\alpha, \delta)=(17^{h} 54^{m} 25^{s}.02, -28^{\circ} 36{'} 48{''}.8)$ in equatorial coordinates, which corresponds to $(\ell,b)=(1^{\circ}.266,-1^{\circ}.489)$ in Galactic coordinates. 

This event was first discovered by the Optical Gravitational Lensing Experiment \citep[OGLE-IV:][]{udalski15} survey using the Warsaw $1.3$ m telescope with $1.4$ square-degree camera, which is located at the Las Campanas Observatory in Chile. The event was announced by the Early Warning System \citep[EWS:][]{udalski94, udalski03} on July 6, 2019 (19:03 UT). The observed data were reduced using their pipeline which employs difference image analysis \citep[DIA,][]{alard98} as modified by \citet{wozniak00}.

% Table 1 -------------------------------------------------------------------
%\begin{landscape}
\begin{deluxetable*}{lrrrr}
\tablecaption{The Best-fit Parameters of STD and TPRX Models\label{table:models}}
\tablewidth{0pt}
\tablehead{
% ---------------------------------------------------------------------------
\multicolumn{1}{c}{Parameter} &
\multicolumn{1}{c}{STD (PSPL)} &
\multicolumn{1}{c}{STD (FSPL)} &
\multicolumn{1}{c}{TPRX $(u_0 < 0)$} & 
\multicolumn{1}{c}{TPRX $(u_0 > 0)$} 
% ------------------------------------
}
\startdata
% -----------------------------------------------------------------------------------------------------------------------------
$\chi^2 / {\rm dof}$ & $ 8729.376 / 8003      $ & $ 8027.218 / 8002        $ & $ 8002.419 / 8000    $ & $  8002.107 / 8000      $ \\
$t_0$ (${\rm HJD'}$) & $ 8670.5661 \pm 0.0001 $ & $ 8670.5649 \pm 0.0001 $ & $ 8670.5649 \pm 0.0001 $ & $  8670.5650 \pm 0.0001 $ \\
$u_0$                & $   -0.0029 \pm 0.0001 $ & $   -0.0037 \pm 0.0002 $ & $   -0.0038 \pm 0.0002 $ & $     0.0037 \pm 0.0002 $ \\
$t_{\rm E}$ (days)   & $    1.6938 \pm 0.0444 $ & $    2.5329 \pm 0.0965 $ & $    2.4984 \pm 0.0914 $ & $     2.5248 \pm 0.0900 $ \\
$\rho_{\ast}$        &      \nodata             & $    0.0092 \pm 0.0004 $ & $    0.0093 \pm 0.0003 $ & $     0.0092 \pm 0.0003 $ \\
$\pi_{{\rm E},N}$    &      \nodata             &      \nodata             & $   -1.6777 \pm 2.5714 $ & $     1.4705 \pm 2.4170 $ \\
$\pi_{{\rm E},E}$    &      \nodata             &      \nodata             & $    1.5404 \pm 0.9340 $ & $     1.7983 \pm 0.3961 $ \\
\hline                                                                                                                    
$F_{\rm S, OGLE}$    & $    0.2592 \pm 0.0073 $ & $    0.1690 \pm 0.0068 $ & $    0.1715 \pm 0.0066 $ & $     0.1696 \pm 0.0064 $ \\
$F_{\rm B, OGLE}$    & $   -0.0672 \pm 0.0073 $ & $    0.0226 \pm 0.0067 $ & $    0.0201 \pm 0.0065 $ & $     0.0220 \pm 0.0064 $ \\
% -----------------------------------------------------------------------------------------------------------------------------
\enddata
\tablecomments{
${\rm HJD' = HJD - 2450000}$.
The PSPL and FSPL indicate point-source/point-lens and finite-source/point-lens models, respectively.
}
\end{deluxetable*}
%\end{landscape}
% ---------------------------------------------------------------------------

% Table 2 ---------------------------------------------------------------------------------------------------------------------------------
\begin{deluxetable}{lrr}
\tablecaption{Error Re-scaling factors\label{table:error}}
\tablewidth{0pt}
\tablehead{
% -----------------------------------------------------------------------------------------------------------------------------------------
\multicolumn{1}{c}{Observatory}   &
\multicolumn{1}{c}{$k$}  &
\multicolumn{1}{c}{$e_{\rm min}$} 
}
\startdata
% -----------------------------------------------------------------------------------------------------------------------------------------
OGLE & 1.0249 & 0.0001 \\
KMTC & 1.1979 & 0.0001 \\
KMTS & 1.1682 & 0.0001 \\
KMTA & 1.0405 & 0.0001 \\
% -----------------------------------------------------------------------------------------------------------------------------------------
\enddata
%\tablecomments{
%The limb-darkening coefficient for $I-$band ($\Gamma_{\it I}$) is $0.4538$. 
%}
\end{deluxetable}
% -----------------------------------------------------------------------------------------------------------------------------------------

The Korea Microlensing Telescope Network \citep[KMTNet:][]{kim16} independently discovered this event (assigned KMT-2019-BLG-1538) on July 7, 2019 (03:16 UT) using its network of three identical $1.6$ m telescopes with $4$ square-degree cameras. The telescopes of the KMTNet are located at the Cerro Tololo Inter-American Observatory in Chile (KMTC), the South African Astronomical Observatory in South Africa (KMTS), and the Siding Spring Observatory in Australia (KMTA). The well-separated KMTNet locations on the southern hemisphere allow near-continuous observations. In addition, the separation between the observatories provides the opportunity to measure the terrestrial microlens-parallax effect \citep{gould97} for special cases (e.g., extreme high-magnification events and/or disk-lenses). The observations are reduced using DIA packages called pySIS \citep{albrow09} and pyDIA \citep{albrow17, bramich13}.

\section{Analyses of the Light Curve} \label{ss3}

\subsection{Single-Lens/Single-Point-Source Model} \label{ss31}

In Figure \ref{fig:LCs}, we present the observed light curve of OGLE-2019-BLG-1058 with model light curves and their residuals. The faint source ($I\sim20$) of this event was highly magnified. The maximum magnification ($A_{\rm max}$) is about $220$. Thus, this event is classified as a high-magnification event, which is roughly defined as $A_{\rm max} \gtrsim 100$. High-magnification events are one of the most important channels for discovering planet-host systems due to the high chance (almost $100\%$ if the planetary system exists) of detecting planetary anomalies at the peak of the light curve \citep{griest98}. However, there is no planetary anomaly at the peak of the light curve, i.e., the light curve exhibits the smooth and symmetric shape of a single-lens/single-source microlensing event.  

Thus, we start the light curve analysis from a single-lens/single-source model with a point-source (hereafter, PSPL). The PSPL lensing event can be described using three parameters: $(t_{0}, u_{0}, t_{\rm E})$, which are the time at the peak of the light curve ($t_0$), the closest separation between the source and the center of the angular Einstein ring ($u_0$; i.e., impact parameter), and the time for the source to traverse the Einstein radius ($t_{\rm E}$). 

In Table \ref{table:models}, we present the best-fit model parameters. The $\chi^2 / {\rm dof}$ values in Table \ref{table:models} are obtained after an error re-scaling process. This process is essential to quantitatively compare different models using the $\chi^2$ difference ($\Delta\chi^2$). For the error re-scaling, we adopt the method described in \citet{yee12} using the equation: $e_{\rm new} = k\sqrt{e_{\rm old}^2 + e_{\rm min}^2}$ where the $e_{\rm new}$ is the re-scaled error, the $e_{\rm old}$ is the original error, the $k$ is the error re-scaling factor, and the $e_{\min}$ is the systematics term. To determine the re-scaling factors, we construct a cumulative sum of $\chi^{2}$ function by sorting the data points according to magnification: $\sum_{i}^{N} \Delta\chi_{i}^{2}$ where $\Delta\chi_{i}^{2}$ is $\chi^{2}$ contribution of each point and $N$ is the number of points of each dataset. Then, we make $\Delta\chi_{i}^2 \sim 1$ by choosing factors based on the best-fit model having the lowest $\chi^{2}$. In Table \ref{table:error}, we present error re-scaling factors for each dataset.

% Figure 2 -----------------------------------------------------------------------------------------
\begin{figure}[htb!]
\epsscale{1.00}
\plotone{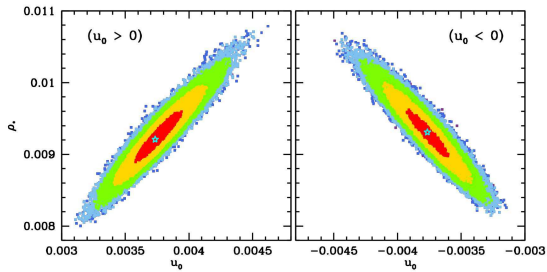}
\caption{
The distributions of the $\rho_{\ast}$ parameter of the TPRX $(u_0 > 0)$ and $(u_0 < 0)$ cases. 
Each color indicates $\Delta\chi^2 \equiv (\chi^2_{\rm chain} - \chi^2_{\rm best-fit}) \le n^2$, 
where $n = 1$ (red), $2$ (yellow), $3$ (green), $4$ (light blue), $5$ (dark blue), $6$ (purple). 
The cyan star indicates the best-fit value of each model.
\label{fig:rho_contour}}
\end{figure}
% --------------------------------------------------------------------------------------------------

From the modeling, we find that the PSPL model cannot describe the peak of the observed light curve (see the zoom-in of Figure \ref{fig:LCs}). Rather, at the peak, the magnification is moderated by the finite-source effect.

\subsection{Single-Lens/Single-Finite-Source Model} \label{ss32}

To describe the shape of the peak, we conduct single-lens/single-source modeling with the finite-source effect (hereafter, FSPL) by incorporating an additional parameter, $\rho_{\ast}$, which is the angular source radius ($\theta_{\ast}$) in units of the angular Einstein ring radius ($\theta_{\rm E}$). That is, $\rho_{\ast} \equiv \theta_{\ast} / \theta_{\rm E}$. For the FSPL modeling, we adopt linear limb-darkening parameters ($\Gamma_{I} = 0.454$ and $\Gamma_{V} = 0.630$) from \citet{claret00} based on the estimated source type (i.e., a late G- or early K-type dwarf source; see Section \ref{ss41}). To check that a linear limb-darkening approximation is sufficiently accurate, we also checked uniform source models. We found a $\chi^{2}$ difference of $< 9$ and no significant difference in the light curve parameters. Because the differences between uniform source and linear limb-darkening profiles are more than an order-of-magnitude larger than the between linear and second-order profiles, we conclude that higher-order limb-darkening models are not necessary.

% Figure 3 -----------------------------------------------------------------------------------------
\begin{figure}[htb!]
\epsscale{1.00}
\plotone{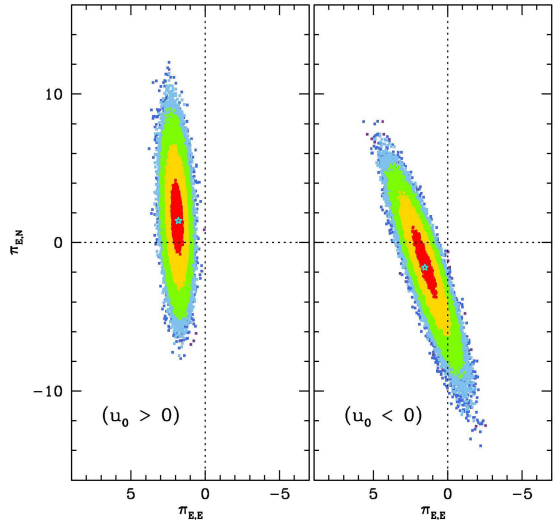}
\caption{
The distributions of the TPRX parameters of the $(u_0 > 0)$ and $(u_0 < 0)$ cases. 
The color scheme is identical to Figure \ref{fig:rho_contour}.
\label{fig:TPRX_contour}}
\end{figure}
% --------------------------------------------------------------------------------------------------

In Table \ref{table:models}, we present the best-fit FSPL model (labeled ``STD (FSPL)"). We find that the FSPL model well describes the peak, which shows a much better fit by $\Delta\chi^2 = 702.158$ than the PSPL model. In Figure \ref{fig:rho_contour}, we present distributions of the $\rho_{\ast}$ parameter, which $\rho_{\ast}$ is well-measured to be $\rho_{\ast} = 0.0092 \pm 0.0004$. We also find that this event lasted for a very short time: $t_{\rm E} = 2.53 \pm 0.10$ days.

\subsection{Terrestrial Microlens-Parallax Model} \label{ss33}

For this event, the timescale is $t_{\rm E} \sim 2.5$ days, which is too short to measure the annual microlens parallax effect \citep{gould92}\footnote[1]{Although the short timescale, we test the annual microlens parallax effect by introducing the microlens-parallax parameters. As expected, we cannot find any $\chi^2$ improvement.}. Also, simultaneous space-based observations have not been taken for this event. Thus, there is no measurement of the space-based microlens parallax effect \citep{refsdal66, gould94}.

Fortunately, this event is a special case: a high-magnification event and, as we will show, plausible with a nearby lens. The characteristics of the event satisfy the conditions for measuring TPRX, e.g., as in the case of OGLE-2007-BLG-224, the source self-crossing time is very short \citep{gould09}. In this case, $t_{\ast} = \rho_{\ast} t_{\rm E} \sim 33$ min. Thus, we test the TPRX effect by introducing parameters of the microlens parallax vector: $\pivec_{\rm E} = (\pi_{{\rm E},{\it N}}, \pi_{{\rm E},{\it E}})$, where $\pi_{{\rm E},{\it N}}$ and $\pi_{{\rm E},{\it E}}$ are components of the microlens parallax vector projected onto the sky along the north and east direction in the equatorial coordinates. The microlens parallax can produce degenerate solutions due to the well-known degeneracy called the ecliptic degeneracy \citep{jiang04, skowron11}. Thus, we test both ($u_0 < 0$) and ($u_0 > 0$) TPRX models. We find $\chi^{2}$ improvements of $24.799$ and $25.111$ for the TPRX ($u_0 < 0$) and ($u_0 > 0$) models (see Table \ref{table:models}). In Figure \ref{fig:TPRX_contour}, we also present distributions of the TPRX parameters. The measured TPRX values of the $(u_0 < 0)$ and $(u_0 > 0)$ solutions are $(\pi_{{\rm E},{\it N}}, \pi_{{\rm E},{\it E}}) = (-1.678 \pm 2.571, 1.540 \pm 0.934)$ and $(1.471 \pm 2.217, 1.798 \pm 0.396)$, respectively. Also, the TRPX distributions clearly exclude $(\pi_{{\rm E},{\it N}}, \pi_{{\rm E},{\it E}}) = (0.0, 0.0)$ suggesting that the effect is clearly detected even though its magnitude is not well-constrained.

% Table 3 ---------------------------------------------------------------------------------------------------------------------------------
\begin{deluxetable}{lrr}
\tablecaption{Properties of the Isolated Lens \label{table:properties}}
\tablewidth{0pt}
\tablehead{
% -----------------------------------------------------------------------------------------------------------------------------------------
\multicolumn{1}{c}{Property} &
\multicolumn{1}{c}{TPRX $(u_0 < 0)$} &
\multicolumn{1}{c}{TPRX $(u_0 > 0)$} 
}
\startdata
% -----------------------------------------------------------------------------------------------------------------------------------------
$\theta_{\rm E}$ (mas)                 & $  0.121  \pm 0.011  $ & $  0.122  \pm 0.011  $ \\
$M_{\rm planet}$ $(M_{\odot})$         & $  0.0065 \pm 0.0058 $ & $  0.0064 \pm 0.0044 $ \\
$M_{\rm planet}$ $(M_{\rm J})$         & $  6.836  \pm 6.027  $ & $  6.740  \pm 4.572  $ \\
$D_{\rm L}$ (kpc)                      & $  2.49   \pm 1.51   $ & $  2.45   \pm 1.16   $ \\
$\mu_{\rm rel}$ $({\rm mas~yr^{-1}})$  & $ 17.691  \pm 1.664  $ & $ 17.603  \pm 1.651  $ \\
% -----------------------------------------------------------------------------------------------------------------------------------------
\enddata
\tablecomments{
These properties were determined using $\theta_{\rm E}$ and $\pi_{\rm E}$ (TPRX) values.
}
\end{deluxetable}
% -----------------------------------------------------------------------------------------------------------------------------------------

\begin{figure}[htb!]
\epsscale{1.00}
\plotone{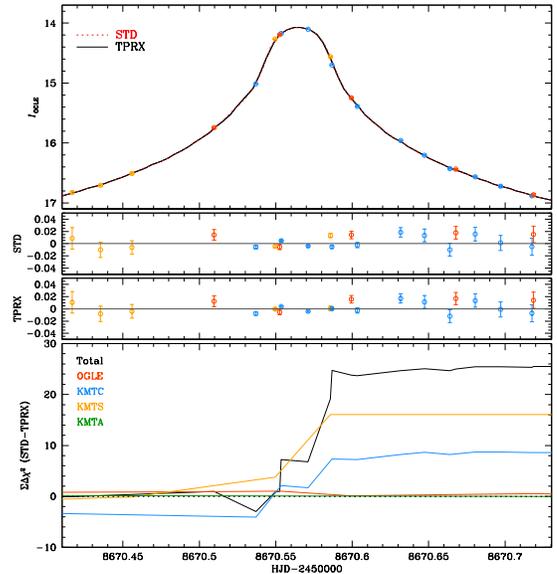}
\caption{
Cumulative $\chi^{2}$ difference ($\Sigma\Delta\chi^{2}$) between STD and TPRX ($u_0 > 0$) models of 
each dataset (lowest panel). We also present the part of the light curve with residuals (upper panels) 
for visual inspection. This part shows $\chi^2$ improvements.
\label{fig:dchi2_TPRX_u0p}}
\end{figure}

However, the TRPX measurements show large uncertainties and the $\Delta\chi^2$ values of both cases are marginal. Thus, we investigate the origin of the $\chi^2$ improvement. In Figure \ref{fig:dchi2_TPRX_u0p}, we present the cumulative $\Delta\chi^2$ plot with the light curve for the TPRX ($u_0 > 0$) case. We find that the $\chi^2$ improvement comes from the peak part of the light curve covered by KMTC and KMTS. This meets our expectations that two or more observatories located in different places should contribute to the TPRX detection and that this effect should be most pronounced over the peak of the event. This expectation is supported by the actual cases of OGLE-2007-BLG-224 and OGLE-2008-BLG-279, which are the two previous examples of TPRX detections \citep{gould09, yee09}.

However, because of the marginal $\Delta\chi^{2}$ improvement, it is not clear whether or not the TPRX measurement is reliable. If we assume the TPRX is correct, we can derive the properties of the lens (see Table \ref{table:properties}). We can also statistically vet the TPRX by comparing the lens properties to the predictions of a Bayesian analysis (without the TPRX constraint). For example, if the TPRX indicated a lens that was a severe outlier relative to the Bayesian prediction, that might indicate the TPRX signal was false or corrupted by systematics.

% Figure 4 -----------------------------------------------------------------------------------------
% --------------------------------------------------------------------------------------------------

\section{Properties of the Lens} \label{ss4}

\subsection{Angular Einstein Ring Radius} \label{ss41}

\begin{figure}[htb!]
\epsscale{1.00}
\plotone{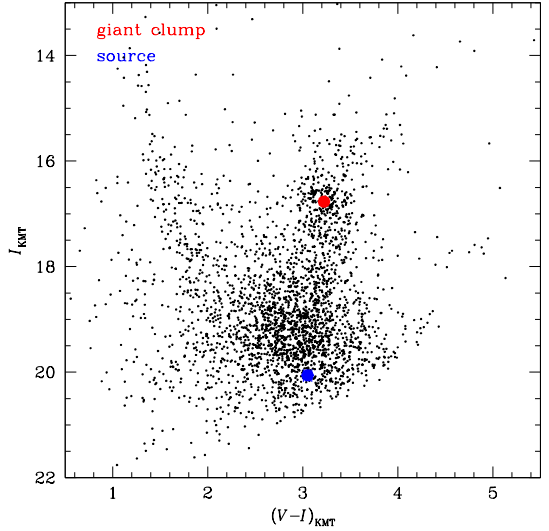}
\caption{
The color-magnitude diagrams constructed using KMTNet observations (KMTC). 
The red and blue dots represent positions of the centroid of the giant clump and the source of the event.
\label{fig:KMTcmd}}
\end{figure}

The determination of the angular Einstein ring radius ($\theta_{\rm E}$) is important because $\theta_{\rm E}$ is a fundamental unit for the analysis of the microlensing event. When the finite-source effect occurs, $\theta_{\rm E}$ can be determined from $\rho_{\ast} = \theta_{\ast} / \theta_{\rm E}$. The angular source size, $\theta_{\ast}$ is an observable, which can be measured from multi-band observations (e.g., $I$- and $V$-bands) using the method of \citet{yoo04}. For this event, there exists a clear signature of the finite-source effect. Thus, it is possible to measure $\rho_{\ast}$. In addition, the KMTNet survey regularly takes $V$-band images for each field. Thus, we can obtain the position of the source on the color-magnitude diagrams (CMD) for the event. By applying Yoo's method, we can obtain the de-reddened source color. In Figure \ref{fig:KMTcmd}, we present the KMT CMD with the source position. For the giant clump centroid, the de-reddened color and magnitude are adopted from \citet{bensby11} and \citet{nataf13}, respectively. Then, we can determine $\theta_{\ast}$ using a color/surface-brightness relation \citep{kervella04}. Using the relation requires first converting the source color from $(V-I)$ to $(V-K)$. For the conversion, we use the color-color relation of \citet{BB88}. The measured color and $I$-magnitude of the source are
\begin{equation}
(V-I, I) = (3.501 \pm 0.045, 20.061 \pm 0.001).
\end{equation}
The de-reddened values are
\begin{equation}
(V-I, I)_{0} = (0.888 \pm 0.067, 17.681 \pm 0.001).
\end{equation}
The determined angular source radius is
\begin{equation}
\theta_{\ast} = 1.121 \pm 0.086 \,{\rm \mu as}.
\end{equation}
By combining this with $\rho_{\ast}$, $\theta_{\rm E}$ is determined:
\begin{equation}
\theta_{\rm E} = 0.122 \pm 0.011 \,{\rm mas}.
\end{equation}
Then, because $\theta_{\rm E}$ is determined, we can determine the lens-source relative proper motion ($\mu_{\rm rel}$), which is 
\begin{equation}
\mu_{\rm rel} = \theta_{\rm E} / t_{\rm E} = 17.58 \pm 1.66 \,{\rm mas\, yr^{-1}}.
\end{equation}
These values for $\theta_{\rm E}$ and $\mu_{\rm rel}$ are derived for the STD solution, but the variation between different solutions is smaller than the uncertainties (see Table \ref{table:properties}).

\subsection{Free-Floating Planet Candidate} \label{ss42}

\subsubsection{Qualitative Analysis} \label{ss421}

The timescale of this event is about $2.5$ days. This unusually short timescale suggests that this single-lens event is a good candidate for a very low-mass object such as a free-floating planet (FFP) or brown dwarf (see Equation (1) and (2)). On the other hand, the measured $\theta_{\rm E}$ is small relative to typical stellar lenses but substantially larger than $\mathcal{O}(10\, {\rm \mu as})$, which is more typical for a strong FFP candidate (see Section \ref{ss1}).

At the same time, the large lens-source relative proper motion, $\mu_{\rm rel} = 17.58 \pm 1.66\, {\rm mas\, yr^{-1}}$ (STD (FSPL) model), suggests that the lens may be in the disk\footnote[2]{Also, note that because $\mu_{\rm rel} > 10\, {\rm mas\, yr^{-1}}$, we cannot use the relation of \citet{kimyh21} to estimate $M_{\rm L}$ from $\theta_{\rm E}$.}. Figure \ref{fig:mu_S_bulge} shows that if the proper motion of the source is typical of bulge stars in this field\footnote[3]{As determined from {\it GAIA} proper motions of bulge clump giants.}, i.e., if $\muvec_{\rm S}(\ell, b) = (-6.232, -0.103)\, {\rm mas\, yr^{-1}}$, $\mu_{\rm rel}$ gives a lens that is inconsistent with having a bulge-like proper motion (the actual source proper motion measurement is discussed in Section \ref{ss43}). However, it would be consistent with belonging to the population of disk stars (whose proper motions are centered near $\muvec_{\rm S}(\ell, b) \sim (0.0, 0.0)\, {\rm mas\, yr^{-1}}$, but with some minor adjustment due to the reflex motion of the Earth and velocity of the Sun with respect to the local standard of rest). If so, that would suggest $\pi_{\rm rel} \gtrsim 0.075\, {\rm mas} $ (for $D_{\rm L} \lesssim 5\, {\rm kpc}$), $\pi_{\rm E} > 0.61$ and thus, $M_{\rm L} \lesssim 0.024 M_{\odot}$. This would be consistent with the constraints on $\pi_{\rm E}$ shown in Figure \ref{fig:TPRX_contour} and would strengthen the case of this lens to be a free-floating planet candidate with a measured TPRX.

% Figure 5 -----------------------------------------------------------------------------------------
% --------------------------------------------------------------------------------------------------

\subsubsection{Bayesian Analysis} \label{ss422}

We expand on this qualitative impression by conducting a full Bayesian analysis using only the $t_{\rm E}$ and $\mu_{\rm rel}$ constraints. To conduct the Bayesian analysis, we construct Galactic priors by adopting several studies. Then, we apply constraints obtained from the characteristics of the event. 

The Galactic prior includes information on mass functions, matter density profiles, and velocity distributions to generate artificial microlensing events. For the mass function, we adopt the multiple-part power-law form of \citet{kroupa02} as the basic form of the mass function. That is,
%\begin{equation}
%\xi(M) = \nu
%	\begin{cases}
%	\left( \frac{M}{M_{1}} \right)^{-\alpha_{0}} & \quad \text{ if } M_{0} < M \leq M_{1} \quad (n=0) \\
%	\left( \frac{M}{M_{1}} \right)^{-\alpha_{1}} & \quad \text{ if } M_{1} < M \leq M_{2} \quad (n=1) \\
%	\left[ \displaystyle\prod_{i=2}^{n\geq2} \left( \frac{M_{i}}{M_{i-1}} \right)^{-\alpha_{i-1}} \right] \left( \frac{M}{M_{n}} \right)^{-\alpha_{n}}        & \quad \text{ if } M_{n} < M < M_{n+1} \quad (n\geq2) \\
%	\end{cases}
%\end{equation}
\begin{equation}
\xi(M) = \nu
	\begin{cases}
	\left( \frac{M}{M_{1}} \right)^{-\alpha_{0}} & \text{(i)}  \\
	\left( \frac{M}{M_{1}} \right)^{-\alpha_{1}} & \text{(ii)} \\
	\left[ \displaystyle\prod_{i=2}^{n\geq2} \left( \frac{M_{i}}{M_{i-1}} \right)^{-\alpha_{i-1}} \right] \left( \frac{M}{M_{n}} \right)^{-\alpha_{n}} & \text{(iii)} \\
	\end{cases}
\end{equation}
\begin{equation}
	\begin{matrix}
	\text{ if (i)~~} &  M_{0} < M \leq M_{1} \,\,~~~~ (n=0) \\
	\text{ if (ii)~} &  M_{1} < M \leq M_{2} \,\,~~~~ (n=1) \\
	\text{ if (iii)} &  M_{n} < M < M_{n+1}  \,~ (n\geq2) \\
	\end{matrix}
\end{equation}
where $\nu = 0.877 \pm 0.045\, {\rm stars}/({\rm pc}^{3} M_{\odot})$ is a scaling factor adopted from \citet{kroupa02}. Then, we also include a component for the FFP mass range: $M \leq 0.013\, M_{\odot}$. The exponents and mass ranges are
\begin{equation}
	\begin{matrix}
	\alpha_{0} = \alpha_{p} & \text{ if } ~~~~~~~~~~\, M/M_{\odot} \leq 0.013 & (n=0)    \\
	\alpha_{1} = 0.3        & \text{ if }   0.013 < M/M_{\odot} \leq 0.080    & (n=1)    \\
	\alpha_{2} = 1.3        & \text{ if }   0.080 < M/M_{\odot} \leq 0.500    & (n=2)    \\
	\alpha_{n} = 2.3        & \text{ if }   0.500 < M/M_{\odot} ~~~~~~~~~~~   & (n\geq3) \\
	\end{matrix}
\end{equation}
where the $\alpha_{p}$ describes the FFP mass function. However, the mass function of FFPs is not well-determined. Thus, we adopt three types of mass functions that have been adopted in previous studies. First, we assume a flat mass function for planet-mass objects: $\alpha_{p} = 1.0$. Second, we adopt the mass function of \citet{zhang20}: $\alpha_{p} = 0.6$. Third, we adopt the mass function of \citet{sumi11}: $\alpha_{p} = 1.3$.

For the matter density profile, we adopt the density profile of the disk from \citet{robin03} and \citet{bennett14}. For the bulge, we adopt the density profile from \citet{hangould95} and \citet{dwek95}. The density profile depends on the distance and location of the objects (i.e., lens and source).

For the mean velocity and velocity dispersion of stars in the bulge, we adopt the {\it GAIA} proper motion information: $\muvec_{\rm bulge}(\ell, b) = (-6.232 \pm 3.304, -0.108 \pm 2.992)\, {\rm mas\, yr^{-1}}$ according to the event's position in Galactic coordinates \citep{gaia18}. For stars in the disk, we adopt the velocity distribution from the refined model of \citet{hangould95}, which is described in \citet{han20}.

We build the Galactic priors using each mass function. Thus, there are three total Galactic-prior sets, one for each FFP mass function. For each set, we generate $4\times10^{7}$ (for each of two cases: bulge or disk lenses) artificial microlensing events using the Monte-Carlo method. Then, we apply constraints on the Galactic priors using weight functions obtained from the models of the event to obtain the posteriors (i.e., the probability distribution of each lens property). 

The first constraint is the timescale ($t_{\rm E}$) of the event. The $t_{\rm E}$ distributions show simple Gaussian distributions. Thus, we can build a weight function using the best-fit $t_{\rm E}$ parameters. That is,
\begin{equation}
W(t_{\rm E}) = \exp \left[ -\frac{1}{2} \left( \frac{t_{\rm E} - t_{\rm E,best}}{\sigma_{ t_{\rm E} } } \right)^{2} \right], 
\end{equation}
where the $t_{\rm E, best}$ and $\sigma_{ t_{\rm E} }$ are the best-fit $t_{\rm E}$ parameters value and their uncertainties, respectively (see Table \ref{table:models}).

The second constraint is the lens-source relative proper motion ($\mu_{\rm rel} = \left|\, \muvec_{\rm L} - \muvec_{\rm S} \right| = \theta_{\rm E} / t_{\rm E}$). Because we have the information of the source proper motion ($\muvec_{\rm S}$), and also the $\mu_{\rm rel}$ quantity already contains the information of the angular Einstein ring radius ($\theta_{\rm E}$), $\mu_{\rm rel}$ can be a stronger constraint than $\theta_{\rm E}$ constraint in this case. Thus, we apply the $\mu_{\rm rel}$ constraints using a Gaussian form. That is,
\begin{equation}
W(\mu_{\rm rel}) = \exp \left[ -\frac{1}{2} \left( \frac{\mu_{\rm rel} - \mu_{\rm rel,best}}{\sigma_{ \mu_{\rm rel} } } \right)^{2} \right], 
\end{equation}
where $\mu_{\rm rel, best}$ and $\sigma_{ \mu_{\rm rel} }$ are determined from the best-fit models. We note that the uncertainty ($\sigma_{ \mu_{\rm rel} }$) is dominated by $\theta_{\ast}$ error. Because $\mu_{\rm rel} = \theta_{\ast} / (t_{\rm E}\, \rho_{\ast})$ and $t_{\rm E}$ and $\rho_{\ast}$ are well-determined, the denominator term is well-defined and its uncertainty is negligible.

The top rows of Figures \ref{fig:bayes_mass} and \ref{fig:bayes_distance} show the results of the Bayesian analysis for the three different FFP mass functions. For this Bayesian analysis, we apply the $t_{\rm E}$ and $\mu_{\rm rel}$ constraints. Contrary to our expectation of an FFP lens in the disk (see Table \ref{table:properties}), these distributions (i.e., distributions in green) show a strong preference for somewhat more massive lenses in the bulge ($M_{\rm L} = 0.08 - 0.09 \, M_{\odot}$, $D_{\rm L} = 7.1 - 7.2$ kpc, see Table \ref{table:bayesian}), although with a tail toward lower-mass disk lenses.

\begin{figure}[htb!]
\epsscale{1.00}
\plotone{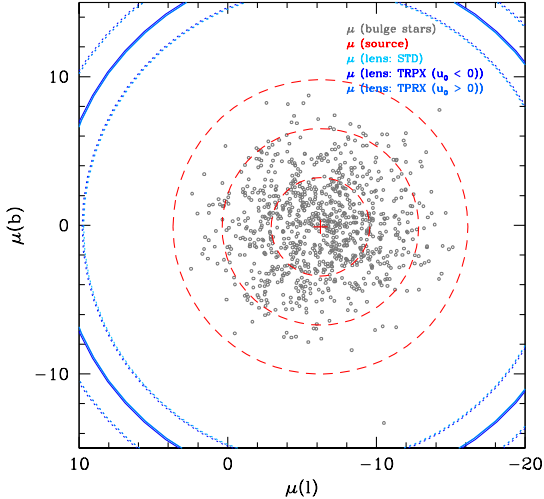}
\caption{
The proper motion diagram. The grey dots present proper motions of red giant clump stars in the 
bulge, which are measured by the {\it GAIA} space telescope. The red cross indicates that the 
mean proper motion of the bulge: $\langle\muvec_{\rm bulge}\rangle = (-6.232 \pm 3.304, -0.103 \pm 2.992)$. 
Dashed circles present $1$, $2$, and $3\sigma$ ranges of the bulge proper motion. Assuming 
$\muvec_{\rm S} = \langle\muvec_{\rm bulge}\rangle$, blue circles (solid lines) present possible positions of 
the lens proper motions with their $1\sigma$ errors (dotted lines), which are yielded from 
the $\mu_{\rm rel}$ values of the STD (cyan) and TPRX (dark and light blue) models.
\label{fig:mu_S_bulge}}
\end{figure}

In fact, this situation is similar to the case of MOA-2011-BLG-262, in which \citet{bennett14} found that bulge-bulge lensing can produce fairly large lens-source proper motions. In the present case of OGLE-2019-BLG-1058, examination of Figure \ref{fig:mu_S_bulge} shows why this is the case. A bulge source with a proper motion equal to the mean motion of the bulge requires a disk lens to achieve $\mu_{\rm rel} = 17.58\, {\rm mas\, yr^{-1}}$. However, because the velocity dispersion of the bulge is relatively large, such a large relative proper motion can also be produced by having a source and a lens a couple $\sigma$ on opposite extremes of the bulge velocity distribution\footnote[4]{This is particularly true in the Bayesian models, which approximate the velocity distribution as Gaussian despite the fact that Figure \ref{fig:mu_S_bulge} demonstrates the true distribution is more centrally concentrated.}. Hence, the Bayesian models prefer bulge-bulge lensing because of the increased density and volume factors despite the lower probability from the velocity distribution.

This suggests that, because of the large lens-source relative proper motion, a measurement of the {\it source} proper motion can provide a strong constraint on the location of the lens. If the measured source proper motion is similar to the mean proper motion of bulge stars, this would strongly favor an FFP lens in the disk (these proper motions correspond to the ``tail" of lower-mass disk lenses). By contrast, if the measured source proper motion is at the edge of the bulge distribution, it would also allow for a larger mass lens in the bulge, which would be preferred over disk lenses by the Bayesian analysis.  

% Figure 6 -----------------------------------------------------------------------------------------
% --------------------------------------------------------------------------------------------------

\subsection{Measured Source Proper Motion} \label{ss43}

In this case, we directly measure the proper motion of the baseline object from the OGLE-IV survey (Udalski et al., in prep). The measured proper motion is $\muvec_{\rm S}(\ell, b) = (-12.46 \pm 2.41, 0.24 \pm 2.41)\, {\rm mas\, yr^{-1}}$ in Galactic coordinates.

The models show that the blend ($F_{\rm B,OGLE}$) makes only a minor contribution to the baseline flux. Specifically, $F_{\rm B,OGLE}/(F_{\rm S,OGLE}+F_{\rm B,OGLE}) = 10 \sim 11\%$ depending on the model. Thus, the baseline object is dominated by light from the source, and therefore, it is reasonable to assume that the proper motion of the baseline object is the proper motion of the source.

\subsubsection{Qualitative Analysis} \label{ss431}

\begin{figure}[htb!]
\epsscale{1.00}
\plotone{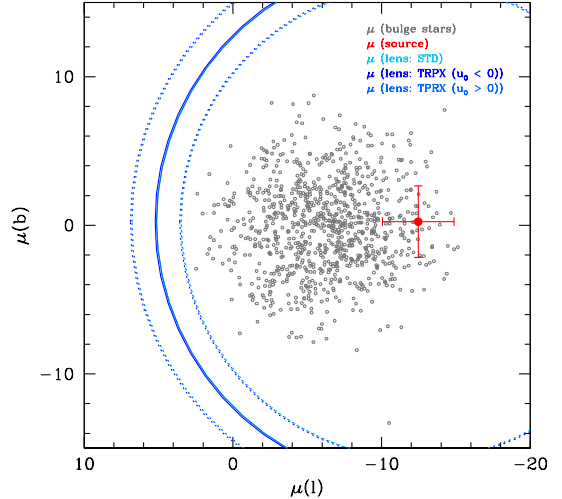}
\caption{
The proper motion diagram with measured $\muvec_{\rm S} = (-12.46 \pm 2.41, 0.24 \pm 2.41)$ (red). 
As in Figure \ref{fig:mu_S_bulge}, the gray points show the proper motions of red clump giants 
in this field as measured by the {\it GAIA} space telescope. The blue circles are the same as 
in Figure \ref{fig:mu_S_bulge} except now they are centered on the measured $\muvec_{\rm S}$ 
instead of the mean bulge value.
\label{fig:mu_S_measured}}
\end{figure}

Under this assumption, the proper motion places the source at the edge of the bulge proper motion distribution (Figure \ref{fig:mu_S_measured}). The blue circles in Figure \ref{fig:mu_S_measured} indicate the range of possible lens proper motions for such a source, given $\mu_{\rm rel}$. Thus, if the source proper motion is the same as the proper motion of the baseline object, the lens would still be consistent with being a disk star. However, there also exist bulge stars within the range of allowed lens proper motions. So, the source proper motion does not rule out bulge lenses, as it would if its proper motion were more typical. However, a bulge lens would imply $\pi_{\rm E} \lesssim 0.34$, which is in tension with the constraints on $\pi_{\rm E}$ from the TPRX.

% Figure 7 -----------------------------------------------------------------------------------------
% --------------------------------------------------------------------------------------------------

\subsubsection{Bayesian Analysis} \label{ss432}

We incorporate the source proper motion measurement into the Bayesian analysis by using $\muvec_{\rm S}(\ell, b) = (-12.46 \pm 2.41, 0.24 \pm 2.41)\, {\rm mas\, yr^{-1}}$ to define the velocity distribution for the source. The results are shown in the third row of Figures \ref{fig:bayes_mass} and \ref{fig:bayes_distance} (i.e., distributions in purple). The result is very similar to the case with only $t_{\rm E}$ and $\mu_{\rm rel}$ information (top row, Section \ref{ss422}) except that lenses with $D_{\rm L} < 2$ kpc have been suppressed. This case also indicates that the lens is a low-mass star in the bulge, whose properties are almost identical to the above case. In Table \ref{table:bayesian}, we present the median value and $68\%$ confidence intervals (i.e., $1\sigma$ errors) of each case.  

\subsubsection{Hypothetical Measurement} \label{ss433}

% Figure 8 -----------------------------------------------------------------------------------------
\begin{figure}[htb!]
\epsscale{1.20}
\plotone{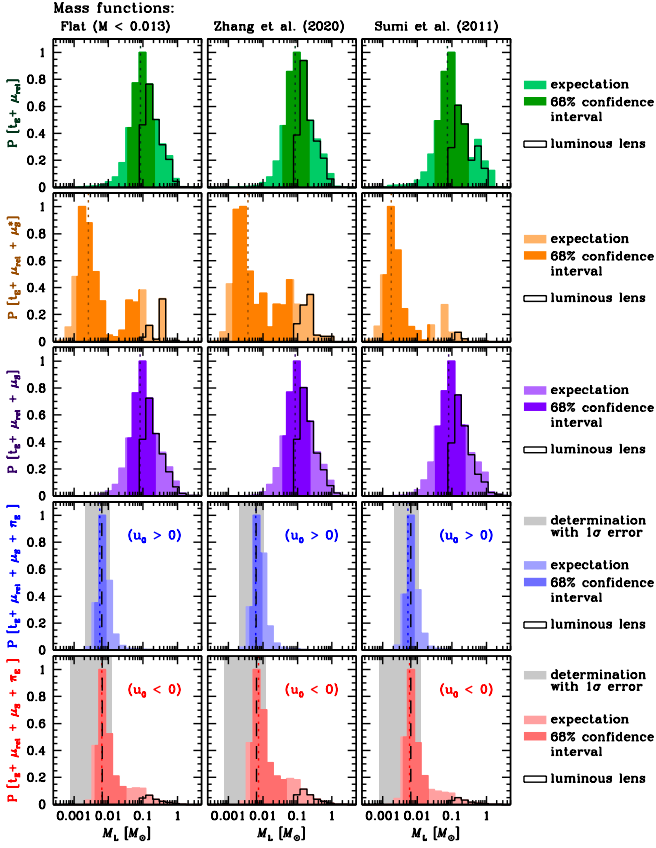}
\caption{
Posterior distributions of the lens mass ($M_{\rm L}$) applying various combinations of constraints. 
The column indicates each mass function used for priors. Green, orange, and purple distributions are 
obtained using constraints of the event timescale ($t_{\rm E}$) and the lens-source relative proper 
motion ($\mu_{\rm rel}$). These use different constraints for the source proper motions 
($\muvec_{\rm S}$): $\muvec_{\rm S}$ drawn from the {\it GAIA} proper motion distribution (green), 
a hypothetical $\muvec_{\rm S}^{\ast}$ with the mean value for the bulge but 
$\sigma_{\mu} = 0.1\, {\rm mas\, yr^{-1}}$ in each direction (orange), and the measured $\muvec_{\rm S}$ 
(purple, blue, and pink). Blue and pink distributions are obtained using constraints of $t_{\rm E}$, 
$\mu_{\rm rel}$, and TPRX ($u_0 > 0$ (blue) and $u_0 < 0$ (pink) cases). The dotted line indicates 
the median value of each distribution. The dashed line and grey shade indicate the property and 
uncertainty, which are determined using the TPRX and $\theta_{\rm E}$ information. The black solid 
line distributions show the luminous lens distributions. Thus, {\it if} $\muvec_{\rm S}$ had 
been measured to be similar to the mean bulge motion, this would have supported the TPRX detection 
(orange vs. red/blue distributions). However, the measured $\muvec_{\rm S}$ is in tension with the 
TPRX (purple vs. red/blue distributions), which suggests that the TPRX measurement may be spurious.
\label{fig:bayes_mass}}
\end{figure}
% --------------------------------------------------------------------------------------------------

In this case, the measured source proper motion did not provide additional evidence to support the conclusion that the lens was an FFP candidate in the disk. However, to show the potential strength of this test, we also conduct the Bayesian analysis using a hypothetical $\muvec_{\rm S}$ measurement. If $\muvec_{\rm S}$ is accurately measured and the source belongs to the bulge, we expect that the Bayesian result should favor a disk-lens. This is very similar to the case presented in Section \ref{ss422} in which the the source is drawn from the bulge velocity distribution with no constraint on $\muvec_{\rm S}$. That case shows that in order to be constraining, the measurement of $\muvec_{\rm S}$ (if it is typical of bulge stars) will need to be significantly more precise than the typical dispersion for bulge stars. The measurement error in $\muvec_{\rm S}$ observed for this event ($\sigma = 2.41\, {\rm mas\, yr^{-1}}$) is probably not sufficiently precise. However, the source in OGLE-2019-BLG-1058 is very faint; significantly more precise proper motions can be obtained for brighter stars (e.g., for OGLE-2016-BLG-1928, which had $I_{\rm S} \sim 17$, $\muvec_{\rm S}$ was measured with a precision of $\pm 1.0\, {\rm mas\, yr^{-1}}$; \citealt{mroz20b}). In addition, if need be, proper motions could be determined to better precision with Keck or HST for cases of special interest (e.g. \citealt{batista15}, \citealt{bennett15}). Thus, we conduct the Bayesian analysis using a hypothetical measurement: $\muvec_{\rm S}^{\ast}(\ell, b) = (-6.232 \pm 0.1, -0.103 \pm 0.1)\, {\rm mas\, yr^{-1}}$.

% Figure 9 -----------------------------------------------------------------------------------------
\begin{figure}[htb!]
\epsscale{1.20}
\plotone{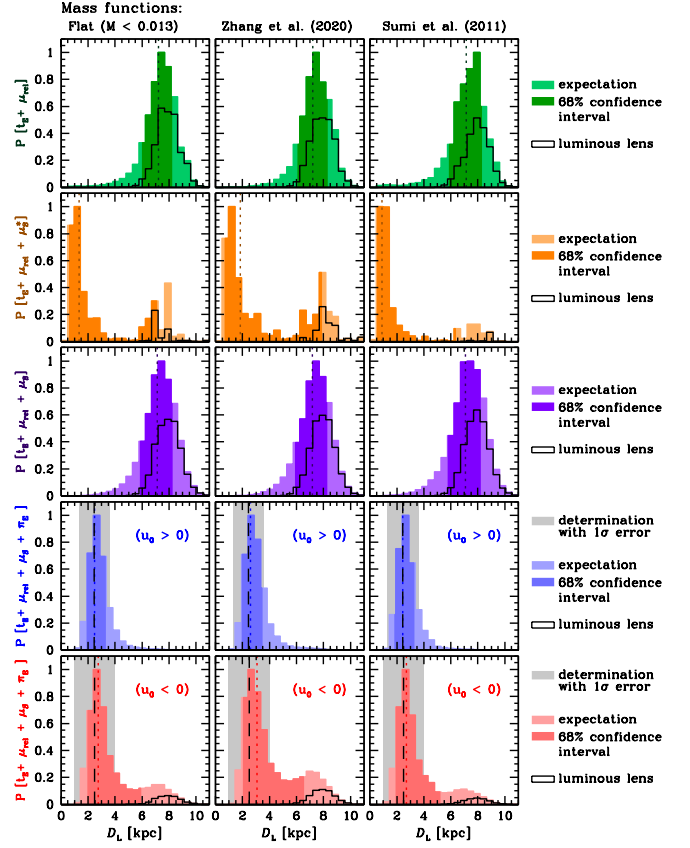}
\caption{
Posterior distributions of the distance to the lens ($D_{\rm L}$). 
The figure scheme is identical to Figure \ref{fig:bayes_mass}.
\label{fig:bayes_distance}}
\end{figure}
% --------------------------------------------------------------------------------------------------

The second rows of Figures \ref{fig:bayes_mass} and \ref{fig:bayes_distance} (i.e., distributions in orange) present the results of this hypothetical case for different mass functions. This result shows that the value of $\muvec_{\rm S}$ and the accuracy with which it is measured both significantly affect the Bayesian test. In cases with large lens-source relative proper motions, if we can show with excellent precision that the source proper motion is typical of bulge stars, the Bayesian analysis would favor a lens in the disk, and consequently, a lens with a very low mass. Thus, such a measurement would provide strong evidence to support the FFP conclusion although higher-mass bulge lenses cannot be completely ruled out.

\subsection{Bayesian Analysis Including TPRX} \label{ss44}

Lastly, we also conduct a Bayesian analysis including TPRX constraints. For the TPRX constraint, we adopt 2D TPRX distributions (i.e., ($\pi_{{\rm E},E}$, $\pi_{{\rm E},N}$) contours) using the method described in \citet{ryu19}. That is,
%\begin{equation}
%W(\pi_{\rm E}) = {\exp} \left[ -\frac{1}{2} \displaystyle\sum^{2}_{i=1} \displaystyle\sum^{2}_{j=1} C_{ij} \left( \pi_{{\rm E},i} -\pi_{{\rm E},i,{\rm best}} \right) \left( \pi_{{\rm E},j} -\pi_{{\rm E},j,{\rm best}} \right) \right]
%~  
%\begin{cases}
%\text{if } i,j = 1 \rightarrow i,j \equiv E \\
%\text{if } i,j = 2 \rightarrow i,j \equiv N
%\end{cases},
%\end{equation}
\begin{equation}
W(\pi_{\rm E}) = {\exp} [ -\frac{1}{2} \displaystyle\sum^{2}_{i=1} \displaystyle\sum^{2}_{j=1} C_{ij} \left( \pi_{{\rm E},i} -\pi_{{\rm E},i,{\rm best}} \right) \nonumber 
\end{equation}
\begin{equation}
\left( \pi_{{\rm E},j} -\pi_{{\rm E},j,{\rm best}} \right) ] \nonumber
\end{equation}
\begin{equation}
\begin{cases}
\text{if } i,j = 1 \rightarrow i,j \equiv E \\
\text{if } i,j = 2 \rightarrow i,j \equiv N
\end{cases},
\end{equation}
where $\pi_{{\rm E},i,{\rm best}}$ and $\pi_{{\rm E},j,{\rm best}}$ are best-fit TPRX parameters. The $C_{ij}$ is the inverse covariance matrix of the TPRX, which describes the shape of TPRX distributions. Then, we apply the TPRX constraints of ($u_0 > 0$) and ($u_0 < 0$) solutions on Galactic priors including the $\muvec_{\rm S}$ measurement. 

% Table 4 ---------------------------------------------------------------------------------------------------------------------------------
\begin{deluxetable}{lrr}
\tablecaption{Bayesian results of the Lens Properties\label{table:bayesian}}
\tablewidth{0pt}
\tablehead{
% -----------------------------------------------------------------------------------------------------------------------------------------
\multicolumn{1}{c}{Case} &
\multicolumn{1}{c}{$M_{\rm planet}$ $(M_{\odot})$} &
\multicolumn{1}{c}{$D_{\rm L}$ (kpc)} 
}
\startdata
% -----------------------------------------------------------------------------------------------------------------------------------------
MF: Flat                                                                    &                           &                     \\
$[t_{\rm E} + \mu_{\rm rel}]$                                               & $0.084_{-0.047}^{+0.127}$ & $7.2_{-1.2}^{+1.0}$ \\
$[t_{\rm E} + \mu_{\rm rel} + \muvec_{\rm S}^{\ast}]$                       & $0.003_{-0.001}^{+0.075}$ & $1.3_{-0.7}^{+5.7}$ \\
$[t_{\rm E} + \mu_{\rm rel} + \muvec_{\rm S}]$                              & $0.083_{-0.046}^{+0.144}$ & $7.1_{-1.2}^{+1.1}$ \\
$[t_{\rm E} + \mu_{\rm rel} + \muvec_{\rm S} + \pivec_{\rm E}]$ $(u_0 > 0)$ & $0.006_{-0.002}^{+0.003}$ & $2.5_{-0.5}^{+0.8}$ \\
$[t_{\rm E} + \mu_{\rm rel} + \muvec_{\rm S} + \pivec_{\rm E}]$ $(u_0 < 0)$ & $0.006_{-0.002}^{+0.026}$ & $2.7_{-0.8}^{+3.2}$ \\
\hline
MF: Zhang+20                                                                &                           &                     \\
$[t_{\rm E} + \mu_{\rm rel}]$                                               & $0.086_{-0.048}^{+0.128}$ & $7.2_{-1.1}^{+1.1}$ \\
$[t_{\rm E} + \mu_{\rm rel} + \muvec_{\rm S}^{\ast}]$                       & $0.004_{-0.002}^{+0.066}$ & $1.8_{-1.1}^{+6.0}$ \\
$[t_{\rm E} + \mu_{\rm rel} + \muvec_{\rm S}]$                              & $0.086_{-0.048}^{+0.142}$ & $7.2_{-1.2}^{+1.1}$ \\
$[t_{\rm E} + \mu_{\rm rel} + \muvec_{\rm S} + \pivec_{\rm E}]$ $(u_0 > 0)$ & $0.006_{-0.002}^{+0.003}$ & $2.6_{-0.6}^{+0.8}$ \\
$[t_{\rm E} + \mu_{\rm rel} + \muvec_{\rm S} + \pivec_{\rm E}]$ $(u_0 < 0)$ & $0.008_{-0.003}^{+0.040}$ & $3.1_{-1.0}^{+3.4}$ \\
\hline
MF: Sumi+11                                                                 &                           &                     \\
$[t_{\rm E} + \mu_{\rm rel}]$                                               & $0.075_{-0.042}^{+0.227}$ & $7.1_{-1.2}^{+1.1}$ \\
$[t_{\rm E} + \mu_{\rm rel} + \muvec_{\rm S}^{\ast}]$                       & $0.002_{-0.001}^{+0.021}$ & $0.9_{-0.4}^{+5.4}$ \\
$[t_{\rm E} + \mu_{\rm rel} + \muvec_{\rm S}]$                              & $0.082_{-0.048}^{+0.161}$ & $7.1_{-1.3}^{+1.1}$ \\
$[t_{\rm E} + \mu_{\rm rel} + \muvec_{\rm S} + \pivec_{\rm E}]$ $(u_0 > 0)$ & $0.005_{-0.002}^{+0.003}$ & $2.4_{-0.6}^{+0.8}$ \\
$[t_{\rm E} + \mu_{\rm rel} + \muvec_{\rm S} + \pivec_{\rm E}]$ $(u_0 < 0)$ & $0.006_{-0.002}^{+0.015}$ & $2.7_{-0.8}^{+2.5}$ \\
% -----------------------------------------------------------------------------------------------------------------------------------------
\enddata
\tablecomments{
$t_{\rm E}$, $\mu_{\rm rel}$, and $\pivec_{\rm E}$ present combinations of constraints.
$\muvec_{\rm S}$ and $\muvec_{\rm S}^{\ast}$ indicate the source proper motion, which is used for Galactic prior.  
The $\muvec_{\rm S}^{\ast}$ is the hypothetical case. For the other cases, the $\muvec_{\rm S}$ is the measured value.
}
\end{deluxetable}
% -----------------------------------------------------------------------------------------------------------------------------------------

In 4th and 5th rows of Figures \ref{fig:bayes_mass} and \ref{fig:bayes_distance} (i.e., distributions in blue and red), we present the Bayesian results of TPRX ($u_0 > 0$) and ($u_0 < 0$) solutions, respectively. The grey shade indicates $M_{\rm L}$ and $D_{\rm L}$ determined from a simple calculation using the $\pi_{\rm E}$ and $\theta_{\rm E}$ measurements (see Equation (3)). We find that the determined properties of both solutions (even though these have large uncertainties caused by the errors of the $\pivec_{\rm E}$) are well-matched with these Bayesian results including the TPRX constraint. Specifically, for the ($u_0 > 0$) case, the expectations and determinations are matched within their $1\sigma$ errors. For the ($u_0 < 0$) case, the median values of the Bayesian posterior match the $1\sigma$ ranges of the determination. But, the $1\sigma$ ranges of the posteriors are wider than the ranges of the determination that derived from simple error propagation. The matched result indicates that the lens would be an FFP located in the disk. In Table \ref{table:bayesian}, we present the values for comparison.

This Bayesian result including the TPRX constraint would be very consistent with the the hypothetical Bayesian results for a case in which $\muvec_{\rm S}$ was measured to be typical for a bulge star (compare bottom rows of Figures \ref{fig:bayes_mass} and \ref{fig:bayes_distance} with the second row). This further demonstrates that such a measurement of the source proper motion would have provided supporting evidence that the TPRX was correct. Of course, the true $\muvec_{\rm S}$ measurement shows that it is the {\it source} proper motion that is unusual.  

Hence, the tension between the results with the TPRX constraint and the results with only the $t_{\rm E}$, $\mu_{\rm rel}$, and $\muvec_{\rm S}$ constraints indicates that the TPRX is highly suspicious. Because of the known possibility that systematics could affect the TPRX (especially, at the $\Delta\chi^2 \sim 25$ level), we adopt the results from the Bayesian analysis including only the $t_{\rm E}$, $\mu_{\rm rel}$, and $\muvec_{\rm S}$ constraints as our fiducial values for the lens properties.  

\section{Test of the Host Existence} \label{ss5}

If the TPRX is correct, the lens is an FFP candidate. For such candidates, we should check whether or not the lens is isolated. This test for a host's existence is essential to confirm an FFP candidate is, in fact, free-floating. Thus, we construct a binary-lens/single-source (2L1S) model by introducing additional parameters describing an additional lens (i.e., a host star). These parameters are the projected separation between binary components normalized by the angular Einstein ring radius of the system ($s$), the mass ratio of the binary components ($q \equiv M_{\rm host} / M_{\rm planet}$), and the angle between the binary axis and the source trajectory ($\alpha$). The other parameters are $t_{0}$, $u_{0}$, $t_{\rm E}$, $\rho_{\ast}$, and the TPRX parameters, $\pi_{{\rm E},N}$, and $\pi_{{\rm E},E}$.  

The 2L1S modeling procedure consists of two steps: a grid search followed by refinement of the model. We use an ($s, q$) grid to search for the existence of a host star. For each grid point, we find the other parameters using the $\chi^2$ minimizing method adopting a Markov chain Monte Carlo (MCMC) algorithm \citep{dunkley05}. Then, we refine the 2L1S model to get the best-fit 2L1S model and its $\chi^2$ value.  

% Figure 10 ----------------------------------------------------------------------------------------
\begin{figure}[htb!]
\epsscale{1.00}
\plotone{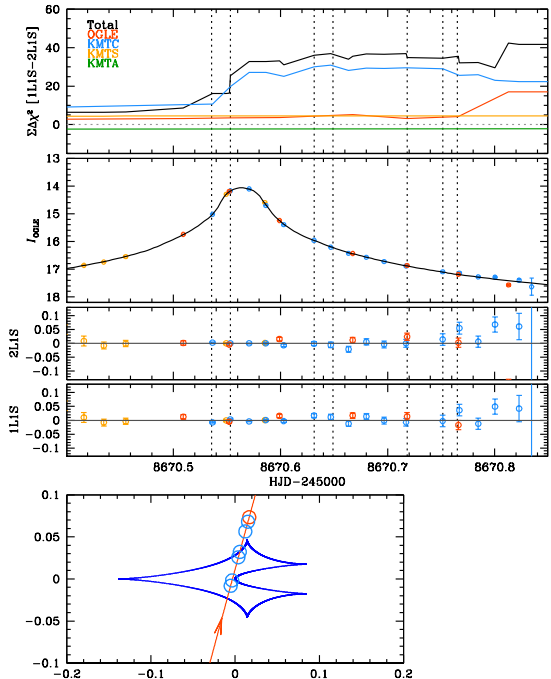}
\caption{
The best-fit 2L1S model light curve of the host test with the cumulative $\Delta\chi^{2}$ plot 
of each dataset. We present the part of the light curve showing the $\chi^2$ improvements. We 
also present the caustic of the 2L1S TPRX model with the source trajectory. The dashed lines 
indicate caustic-crossing points (empty dots on the source trajectory), which yield the $\chi^2$ 
improvements. 
\label{fig:host_test}}
\end{figure}
% --------------------------------------------------------------------------------------------------

The best-fit minimum is at $s\sim1$ with $\log q \sim 3$, which would be consistent with a host star for a planetary lens. We find that $\Delta\chi^{2} \sim 41.24$ between FSPL+TPRX and 2L1S models. We investigate the origin of the $\chi^2$ improvement using the cumulative $\Delta\chi^{2}$ plot of each dataset. We find that the $\chi^2$ mostly improves during the magnified part of the light curve (see Figure \ref{fig:host_test}). Figure \ref{fig:host_test} also, shows the caustic geometry. It shows the $\chi^2$ improvement comes from KMTC and OGLE data, which are fit by extremely subtle deviations ($\le 0.02$ mag) in the light curve when the source crosses complex features of the caustic (see the geometry). This improvement is not reliable because it comes from extremely subtle deviations of only a few data points, and it is not clear that systematics in the photometry are controlled well enough at this level. Hence, there is not enough information to comment on whether or not this lens has a companion. If the lens is later shown to be an FFP candidate rather than a star (see Section \ref{ss6}), these issues could be revisited in more detail.

\section{Discussion and Conclusion} \label{ss6}

We have demonstrated that a measurement of $\muvec_{\rm S}$ is a powerful test of terrestrial microlens parallax (TPRX) signals. In the case of OGLE-2019-BLG-1058, we measured TPRX, but at a significance low enough that it could be affected by systematics in the photometry, and therefore, could not be considered robust by itself. The large $\mu_{\rm rel} = 17.6 \pm 1.7\, {\rm mas\, yr^{-1}}$ appears to support the TPRX signal because both are consistent with a nearby disk lens, which would then be an FFP with $M_{\rm L} \sim 6\, M_{\rm Jupiter}$. However, the OGLE-IV survey gives a measurement of the proper motion of the baseline object, which is dominated by light from the source star. This proper motion measurement ($\muvec_{\rm S} = (-12.46, 0.24)\ {\rm mas\, yr^{-1}}$) is extreme compared to typical bulge stars. Hence, once combined with the $t_{\rm E}$ and $\theta_{\rm E}$ constraints in a Bayesian analysis, the addition of the source proper motion constraint favors a low-mass star ($M_{\rm L} = 0.08 - 0.09\, M_{\odot}$) located in the bulge ($D_{\rm L} = 7.1 - 7.2$ kpc).

Of course, a Bayesian inference is not a measurement. The large $\mu_{\rm rel}$ of $17.6\, {\rm mas\, yr^{-1}}$ implies that the source and lens (i.e., FFP candidate) will be separated enough to attempt high-resolution follow-up observations relatively soon. About $50\%$ of the Bayesian probability implies a stellar, and therefore luminous, lens. Given the measurement of $\theta_{\rm E}$, a stellar lens would also be relatively distant, so behind the same amount of extinction as the source \citep[i.e. $A_{H} = 0.502$;][]{gonzalez12}. Thus, such a lens with $M \gtrsim 0.1\, M_{\odot}$ would have $H \lesssim 26$ \citep{pecaut13}. Based on the source $(V-I)_{0}$ color and $A_{H}$, we estimate the observed source brightness to be $H_{\rm S} = 17.16$ \citep{BB88}. Using the NIRC2 coronagraph, \citet{bowler15} were able to achieve $\sim 5$ magnitudes of contrast in $H$-band at $\sim 100 - 300\, {\rm mas}$ separations and $\sim 10$ magnitudes of contrast outside of that. Scaling to a $30$ m telescope implies that the lens will be far enough from the source to be separately resolved at first light of EELTs, if one had a similar instrument. Hence, such observations could conclusively demonstrate that the lens is a star.  

We also showed that if the source proper motion had been measured (but with better precision) to be in the center of the bulge distribution, our inference about the lens would have been the opposite. Specifically, the Bayesian analysis would have shown a preference for an FFP lens at $\sim 1$ kpc, which would have been supporting evidence that the terrestrial parallax measurement was correct. Hence, this event demonstrates how source proper motions can be used to validate (or invalidate) weak terrestrial parallax measurements, which \citet{gould13} showed can be used to distinguish between true free-floating planets and stellar contaminants. This is particularly important for events with larger Einstein radii ($\gtrsim 10\, \mu$as), i.e., above the Einstein desert posited by \citet{ryu21}.

Finally, we note that this case is similar to the case of OGLE-2017-BLG-0560, which also had a large lens-source relative proper motion \citep[$\mu_{\rm rel} = 15.6\, {\rm mas\, yr^{-1}}$;][]{mroz19}. As in OGLE-2019-BLG-1058, the measured source proper motion was found to be at the edge of the bulge distribution. Hence, \citet{mroz19} reported a relatively large mass ($M_{\rm L} = 100\, M_{\rm Jup}$) and small relative parallax ($\pi_{\rm rel} = 0.002$ mas), which implies a lens in the bulge. By contrast, if they had instead measured a more typical bulge proper motion, there would probably be a case for the lens to be a much lower-mass FFP in the disk.

% =================================
\begin{acknowledgments}
% KMTNet
This research has made use of the KMTNet system operated by the Korea Astronomy and Space Science Institute (KASI), and the data were obtained at three host sites of CTIO in Chile, SAAO in South Africa, and SSO in Australia.  
% Andy
Work by A.G. was supported by JPL grant 1500811.
% Han
Work by C.H. was supported by grants of the National Research Foundation of Korea (2017R1A4A1015178 and 2019R1A2C2085965).
\end{acknowledgments}
% =================================

\end{document}